# Philosophical Survey of Passwords

M Atif Qureshi[1], Arjumand Younus[2] and Arslan Ahmed Khan[3]

[1] UNHP Research
Karachi, Sind, Pakistan
*atif.qureshi@unhp.com.pk*

[2] UNHP Research
Karachi, Sind, Pakistan
*arjumand.younus@unhp.com.pk*

[3] UNHP Research
Karachi, Sind, Pakistan
*arslan.khan@unhp.com.pk*

**Abstract**
Over the years security experts in the field of Information Technology have had a tough time in making passwords secure. This paper studies and takes a careful look at this issue from the angle of philosophy and cognitive science. We have studied the process of passwords to rank its strengths and weaknesses in order to establish a quality metric for passwords. Finally we related the process to human senses which enables us to propose a constitutional scheme for the process of password. The basic proposition is to exploit relationship between human senses and password to ensure improvement in authentication while keeping it an enjoyable activity.

## 1. Introduction

No doubt information is a valuable asset in this digital age. Due to the critical nature of information, be it personal information on someone's personal computer or information systems of large organizations, security is a major concern. There are three aspects of computer security: authentication, authorization and encryption. The first and most important of these layers is authentication and it is at this layer that passwords play a significant role.

Most common authentication mechanisms include use of an alphanumeric based word that only the user to be authenticated knows and is commonly referred to as passwords [1]. The SANS Institute indicates that weak or nonexistent passwords are among the top 10 most critical computer vulnerabilities in homes and businesses [2]. Philosophical analysis of passwords can lead to the refinement of the authentication process. This approach has rarely been adopted in the exploration and design of computer security. Passwords too are entities having an existence of their own and this lead us to study them under a philosophical context.

Passwords: this word is essentially composed of two words i.e. pass and word so you pass if you have the right word. Even before the advent of computers watchwords existed in the form of secret codes, agents of certain command for their respective authorization or administration used watchword e.g. for identifying other agents [3] and the underlying concept is essentially the same today. Next we move on to word: in this context word is not necessarily something making dictionary-based sense (we do keep passwords that make no meaning e.g. passwords like adegj or a2b5et). Hence Passwords are keys that control access. They let you in and keep others out. They provide information control (passwords on documents); access control (passwords to web pages) and authentication (proving that you are who you say you are) [4]. In this paper we take a deep look into both the theory and philosophy of passwords; in short we will be addressing a fundamental question: can password semantics enable them to mimic Nature's way of keeping secrets and providing security.

### 1.1 Why philosophical perspective of passwords

Ontology is a philosophical term used to describe a particular theory about the nature of being or the kinds of things that have existence [5]. In the context of passwords it implies a careful and thorough dive into the existence and nature of passwords and their relationship to users and computers. A password has a relationship with the user's mind and therefore it should be linked with specific user's mindset by creating a sensible bridge between the two. In short password must be backed by a certain philosophy

IJCSI

which establishes a link between concerned rational entities i.e. user and system of recognition.

## 1.2 Outline

The organization of this paper is as follows: in section 2 we take a careful look into the problems of the existing password schemes and analyze the existing solutions. In section 3 we propose some suggestions in light of our philosophical approach at the same time evaluating and presenting a critique of the existing mechanisms. Finally section 4 concludes the discussion.

## 2. The Password Problem

When it comes to the area of computer security there is a heavy reliance on passwords. But the main drawback of passwords is what is termed as the *"password problem"* [6] for text-based passwords. We will refer to this problem as the *"classical password problem."* This problem basically arises from either two of the following facts:

1) Human memory is limited and therefore users cannot remember secure passwords as a result of which they tend to pick passwords that are too short or easy to remember [7]. Hence passwords should be easy to remember.
2) Passwords should be secure, i.e., they should look random and should be hard to guess; they should be changed frequently, and should be different on different accounts of the same user. They should not be written down or stored in plain text. But unfortunately users do not tend to follow these practices [8].

Tradeoffs have to be made between convenience and security due to the shortcomings of text-based passwords. Now we explore some techniques that have been adopted to minimize the tradeoffs and increase computer security.

## 2.1 Attempts to Address the Problem

Current authentication techniques fall into three main areas: token-based authentication, biometric-based authentication and knowledge-based authentication.

Token-based authentication techniques [9] use a mark or a symbol for identification which is only known to the authenticating mechanism and it is under the possession of the user just like a coin which has no meaning other than that known to the mechanism. An example is that of key cards and smart cards. Many token-based authentication systems also use knowledge-based techniques to enhance security. For example ATM cards are generally used together with a PIN number [1]. Biometrics systems are being heavily used [10], biometric authentication refers to technologies that measure and analyze human physical and behavioral characteristics for authentication purposes. Examples of physical characteristics include fingerprints, eye retinas and irises, facial patterns and hand measurements, while examples of mostly behavioral characteristics include signature, gait and typing patterns. Voice is considered a mix of both physical and behavioral characteristics. However, it can be argued that all biometric traits share physical and behavioral aspects.

Knowledge-based techniques are most common and will mainly be the focus of our discussion and under which both text-based and picture-based passwords are subcategorized.

## 2.1 An Extension of Knowledge-Based Passwords

A new phenomenon that computer security researchers have recently explored under the domain of knowledge-based passwords is that of graphical passwords i.e. passwords that are based on pictures. They have motivated their studies on some psychological studies revealing that humans remember pictures better than text [11]. Picture-based passwords are subdivided into recognition-based and recall-based approaches.

Using recognition-based techniques, a user is presented with a set of images and the user passes the authentication by recognizing and identifying the images he or she selected during the registration stage. Using recall-based techniques, a user is asked to reproduce something that he or she created or selected earlier during the registration stage.

## 3. Passwords from a Philosophical Viewpoint

As previously mentioned we focus on an ontological study of passwords and that too under the light of philosophy. However ontology has its definition in Computer Science (more specifically in Artificial Intelligence [5]). In fact at the start of this century emerged a whole new field namely cognitive science [12] which brought scholars of philosophy and computer science close to each other and under this field computer scientists are closely studying working of the human mind to make computational tasks efficient. It is this approach that we also propose and that's one main reason why we say that passwords should be studied from a philosophical perspective.

Passwords have never managed a distinct line whether it is a single unit of work or a process. If the password follows a cognitive paradigm then password recognition is a complete process just like the human mind follows a certain process in recognizing and authenticating known people; similarly computers should take passwords as a process in the light of philosophy. In fact we believe that much of the drawbacks in previous approaches are due to treating password as a unit of work and not carefully viewing the details of the entire process in close context with the human mind. The password recognition process is a detailed DFD (data flow diagram) rather than a context DFD.

Once we are clear that password recognition is a process we must now look at ways that can make this process friendly for the humans at the same time ensuring security to the maximum level. A common point that is raised when addressing the classical password problem defined in the previous section is that human factors are the weakest link in a computer security system [13]. But here we raise an important question: is human really the weak link here or is it the weakness of evaluation procedure for password that under utilizes the intelligence and senses of human that make him look as a naive link in whole process of text-based passwords environment. In fact human intelligence and senses if properly utilized can result in best-possible security mechanism.

### 3.1 Some Problems in Earlier Attempts

In section 2.1 we explored some attempts to solve the classical password problem. However each of the techniques that have been proposed has some drawbacks which can be summarized as follows:

- The token-based passwords though secure but require a token (permit pass) which could be misplaced, stolen, forgotten or duplicated and the biggest drawback is that the technique can only be applied in limited domains not within the reach of common user.

- The biometric passwords are efficient in that they are near to a human's science and do not require remembrance rather they are closely linked with humans but they are expensive solutions and cannot be used in every scenario.

- Knowledge based passwords require remembrance and are sometimes breakable or guessable.

### 3.2 Proposed Directions to Prevent Possible Attacks

Following directions can be adopted in order to improve the security of passwords at the same time making it an enjoyable/sensible activity to ensure user satisfaction:

1. Appropriate utilization of human senses in the passwords.
2. Increase in the domain set of password by introducing a greater deal of variety.
3. Empowering user to make selection from domain set of variety to ensure his mental and physical satisfaction.
4. Introducing facility of randomization into the password.
5. Ensure the establishment of a link between system and specific human mind from domain set.

A discussion on possible attacks and tips for prevention (in light of philosophy and cognitive science) follows:

- **Brute force search:** is basically a global attack on passwords to search for all possible combinations of alpha numerals (in case of text-based passwords) and graphical images (in case of graphical passwords). In short brute force launches attack of words that can be text-based, activity and mixed courses of action. The brute force attack can be prevented with ease by application of point 2, 4 and 5 mentioned above and as a result the brute-force attack becomes computationally impossible. This philosophy should be kept in mind and the engine should be such that point 3 also follows as a logical consequence.

- **Dictionary Attacks:** are regional attacks that run through a possible series of dictionary words, activities and mixed courses of action until one works. Even some graphical passwords are vulnerable to these types of attacks. However these can be prevented in an effective manner by application of techniques mentioned in point 4 and 5. This will allow maximum sense exploitation so dictionary attacks would fail most often.

- **Shoulder surfing:** is when an attacker directly watches a user during login, or when a security camera films a user, or when an electromagnetic pulse scanner monitors the keyboard or the mouse, or when Trojan login screens capture passwords etc [6]. This attack can easily be prevented with the simple approach proposed in point 4 in the pass-word the pass should be the same but we should not take the word as static thereby making it pass-sense.

- **Guessing:** is a very common problem associated with text-based passwords or even graphical passwords. Guess work is possible when the domain is limited and choices are few; in other words there is a lesser utilization of senses. So this threat can easily be prevented by practicing points 1, 3 and 4.

- **Spy ware:** is type of malware that collects user's information about their computational behavior and personal information. This attack can easily fail in the light of above mentioned points 4 and 5 which imply that the password is making sense to both human and computer but not spyware.

All these suggestions were for the knowledge-based passwords but this philosophy can also be applied on other two categories as mentioned in section 3.1. Biometrics and token-based authentication mechanisms cannot be deployed everywhere because of the amount of investment and ease of use. But these authentication mechanisms can be treated as choice for domain set as mentioned in point 2 and leaving the choice to user as discussed in point 3.

### 3.3 Redefinition of the Password Problem

In the classical scenario the domain of the problem was simply limited to text-based passwords but the three solutions proposed: token-based passwords, biometric passwords and knowledge-based passwords (under which come both text-based and graphical passwords) widen the scope of the problem. Furthermore the directions that we have proposed in section 3.2 can lead to other issues in the password arena. The treatment of password recognition as a process and exploitation of human senses in the process seems to be an appealing idea but it naturally leads to a redefinition of the password problem. Hence first of all we must redefine the password problem in order to extend its domain and increase the size of the universe of discourse.

We can redefine the problem as follows:

1. Introducing variety into the domain set of password is a task that must be given due consideration and any attempt to implement the philosophical concepts explored in this paper must address the question: How and in what ways can variety be introduced into the passwords so that $N^K$ formulation sustains more with N than with K where N is single input or action and K is length of input.
2. We have stated that password recognition is a process in itself but the details and phases of that process have to be identified. To accommodate philosophical ideas one must carefully model the process of evaluation (i.e. input and validation).
3. By exploiting senses to ensure variety does not mean to exhaust user both physically and mentally but means to enhance level of comfort and freedom to choose from variety that lead in securing system sensibly.
4. Randomization in password should follow the common sense rather than heavy mental exercise in a way that senses tell computer system "Yes, I am the right person. Please let me pass!"
5. In security critical zones, heavy investment is made to ensure protection at the level of authentication but lacks to decide level of quality achieved. The discussion in section 3.2 will give transparency for proper budgeting, level of comfort and level of security achieved in authentication mechanism.

In short a sensible link between the human mind and the computer system for verification is a complex problem and is a great challenge for researchers in the field of computer security

## 4. Conclusions

This paper has thrown light onto the philosophy of passwords and their study in connection with the human mind. Although the points that were mentioned in this paper have been noted by different researchers at different times but there's no single place where the entire *"password philosophy"* has been defined. Thus we have laid out the constitutional terms for any study of intelligent and smart passwords. The two main points that we have identified in this "*Constitution of Passwords*" are as follows:

1. Password is not just a unit of work; rather it is a complete process.
2. Password should incorporate common sense of humans.
3. There must be quality assurance at the level of authentication mechanism.

This philosophy can play vital role for immediate practitioners if they keep tradeoff of in their mind before producing a secure solution and as well as for researchers to dive into challenging problems that have been left open for them.